# A dual-phase cobalt alloy with a triple yielding phenomenon under compression test

Ali Shafiei

Metallurgy Group, Niroo Research Institute (NRI), Tehran 14665-517, Iran
E-mail: alshafiei@nri.ac.ir, Tel: +98 (21) 88074187

**Abstract**

The compressive mechanical behavior of a dual phase cobalt alloy ($Al_{14}Co_{41}Cr_{16}Fe_{11}Ni_{18}$) is reported in this communication. An uncommon triple yielding phenomenon is observed in the as-cast condition. Microstructural studies suggest that the observed behavior may be due to a stress/strain-induced martensitic phase transformation.

High entropy alloys (HEAs) are founded on the idea that an alloy could have multiple (at least three) principal elements instead of one dominant element [1]. Based on this "fundamentally new idea" [1], new alloys, mostly selected from the central part of the multicomponent phase diagrams, are being designed and investigated; the aim is to identify alloys with enhanced properties in comparison with traditional alloys. Initially, the focus was on equiatomic alloys e. g. CoCrFeMnNi, but currently, non-equiatomic alloys are also being investigated.

The exploration has leaded to the discovery of some HEAs with new physical phenomena and promising properties [1-4].

One of the extensively studied alloy systems in the field of HEAs is Al-Co-Cr-Fe-Ni [3-4]. Depending on the alloy composition, a broad range of microstructures and properties could be obtained. For example, one may name alloy $AlCoCrFeNi_{2.1}$ with an in-situ lamellar composite microstructure showing a good balance of strength and ductility [5]. As another example, one may name alloy $Al_{0.5}CoCrFeNi$ which could be hardened by a precipitation hardening mechanism [6]. In the present work, a Co rich alloy ($Al_{14}Co_{41}Cr_{16}Fe_{11}Ni_{18}$) is introduced which shows a triple yielding phenomenon during compression. The composition was reached during trial and error experiments for finding a eutectic high entropy alloy. Double yield phenomenon is common in alloys with stress/strain-induced martensitic phase transformation (SIM) [7-8], but a triple yielding phenomenon is rare [9]. The objective of this communication is not to characterize the observed behavior, although some speculations are made to explain the mechanisms involved.

Two compressive stress-strain curves for alloy $Al_{14}Co_{41}Cr_{16}Fe_{11}Ni_{18}$ in the as-cast condition are shown in Figure 1. Experimental procedures used for the preparation and testing of samples are explained in [10]. It can be seen that the slope of the compressive stress-strain curves changes at some points indicating the activation of different deformation mechanisms during the compression test. The changes of slope can be better seen in Figure 1b where the points at which changes in slope occur are arrowed.

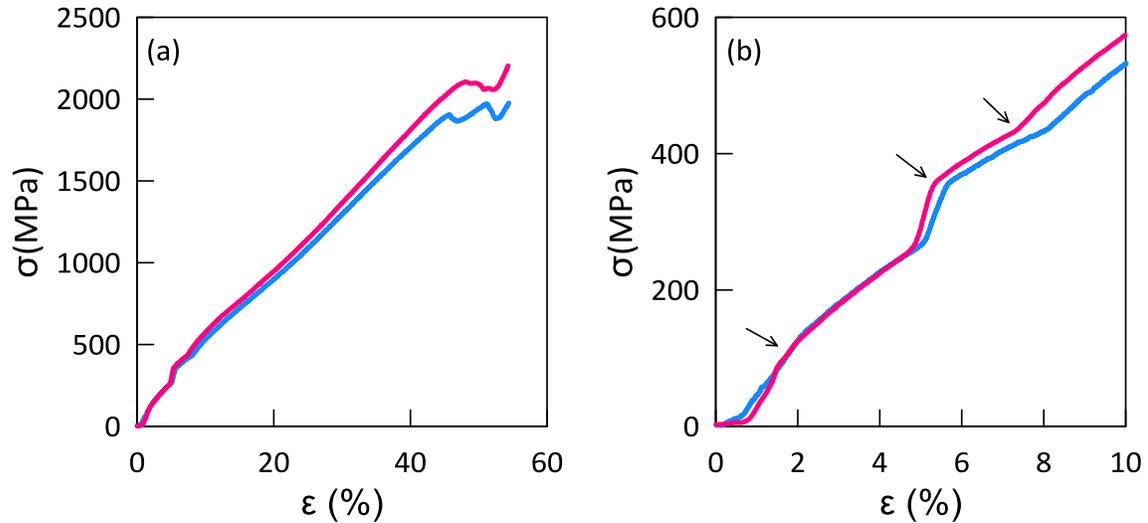

Figure 1. (a) The engineering compressive stress-strain curves for as-cast alloy $Al_{14}Co_{41}Cr_{16}Fe_{11}Ni_{18}$ (b) observing three yielding points at small strains

Double yielding phenomenon is common in alloys with stress/strain-induced martensitic phase transformation (SIM) [7-8]. Therefore, the observed behavior could be attributed to the phase transformations during the compression test. In fact, very recently SIM is reported for some Co-rich HEAs with chemical compositions similar to the chemical composition of alloy investigated here [11-13]. To assess the formation of a martensite phase during deformation, the microstructures of samples before and after compression test were investigated. Optical images of the as cast microstructure are shown in Figure 2. It can be seen that the alloy has a dendrite microstructure and consists of primary dendrites and inter-dendritic regions with a eutectic microstructure. According to the XRD results and thermodynamic simulations [10], the crystal structure of bright and dark regions were determined to be FCC and BCC(B2) respectively. It could be speculated that the deformation during compression test will be mostly confined to the bright regions (including primary dendrites) with a FCC crystal structure because BCC/B2 phase are usually considered to be brittle and not deformable.

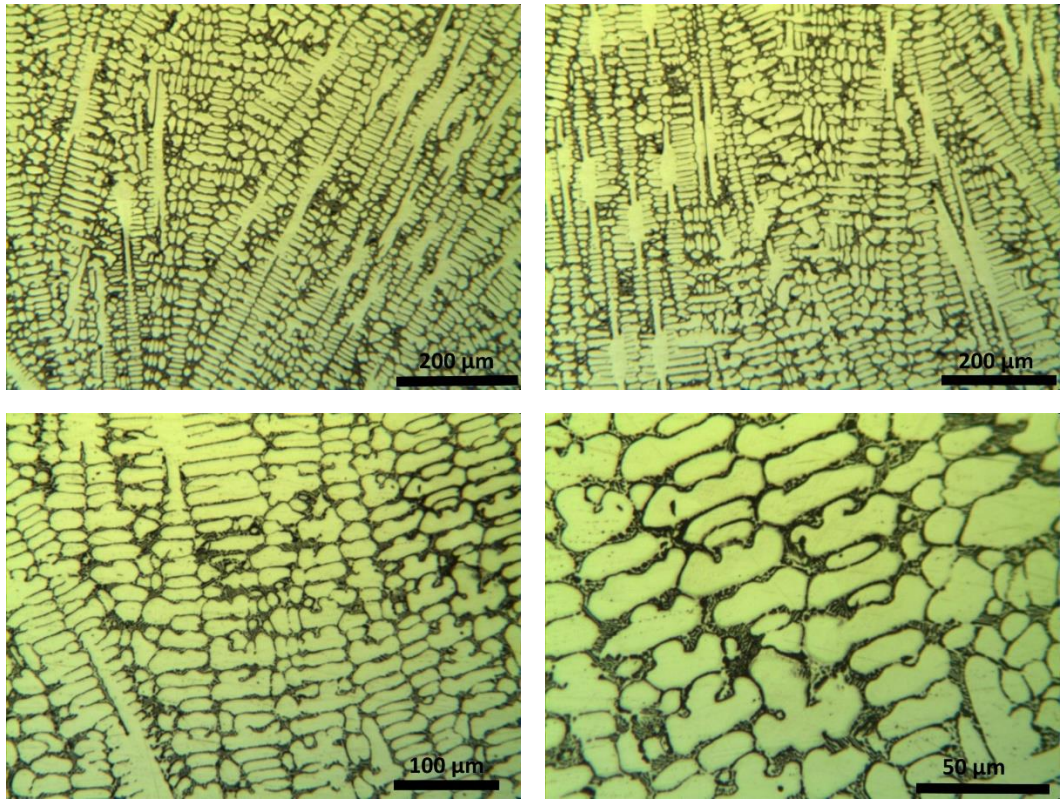

Figure 2. The as cast microstructure of alloy $Al_{14}Co_{41}Cr_{16}Fe_{11}Ni_{18}$

Optical images from the microstructure of one of the deformed samples are shown in Figures 3-5. Images in Figures 3-5 were taken from the surface which was in contact with the die during the compression test. Different microstructures (Figures 3-5) were observed across the investigated surface suggesting that the amount of deformation was not the same across the surface. In some regions, distorted dendrites with no effect of twinning or SIM were observed (Figure 3), while in some regions new phases (probably twins or martensite) were identified inside of dendrites (Figure 4). Some highly deformed areas were also observed

(Figure 5). It seemed that BCC/B2 phases (dark regions) were also undergone some changes in these highly deformed areas (probably cracked during the deformation of dendritic regions).

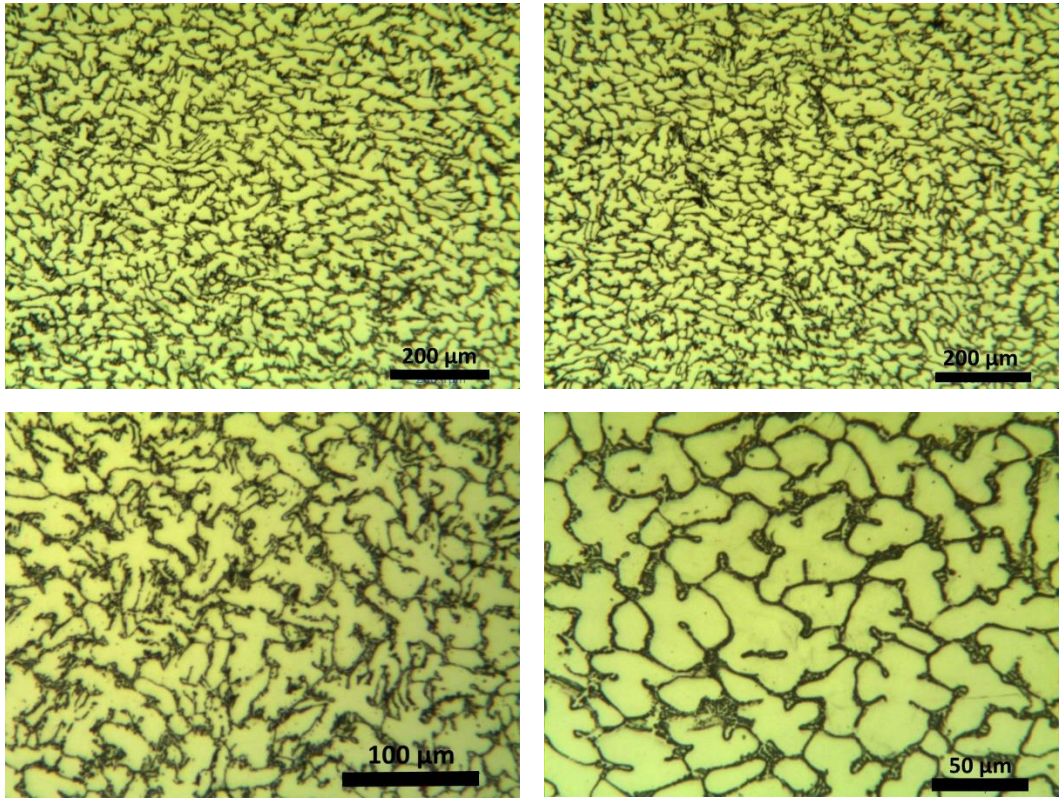

Figure 3. Distorted dendrites after compression test with no effect of twinning or SIM

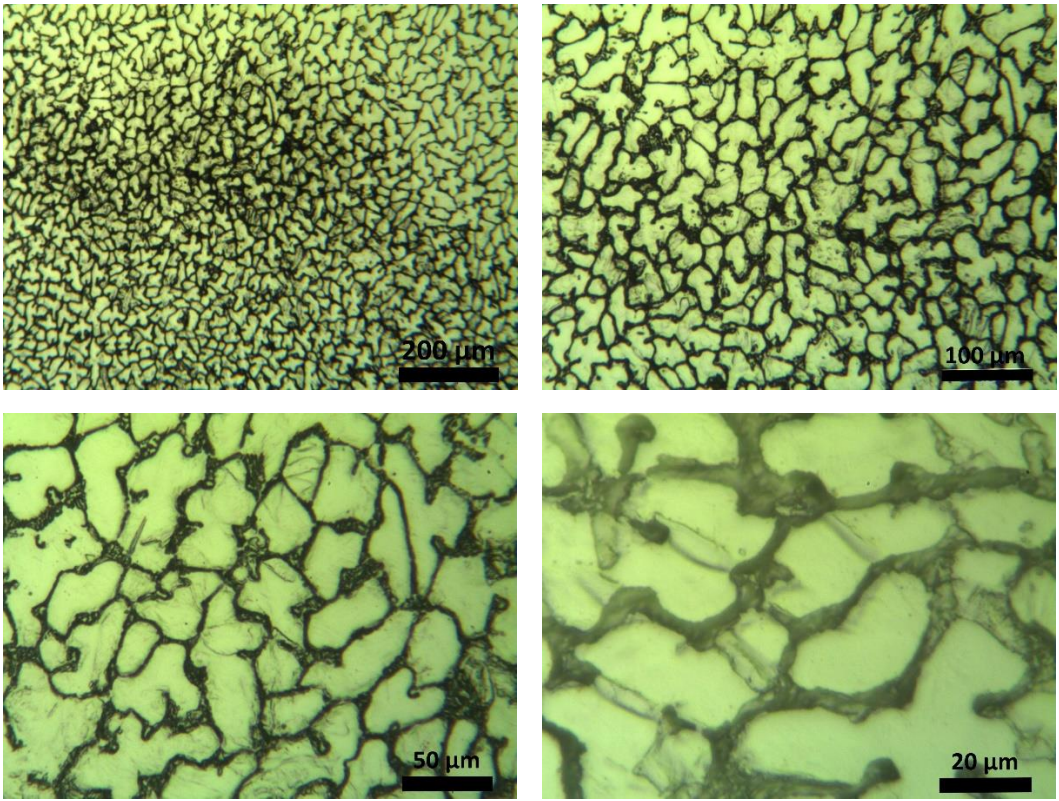

Figure 4. Observation of secondary phases inside of dendrites after compression test

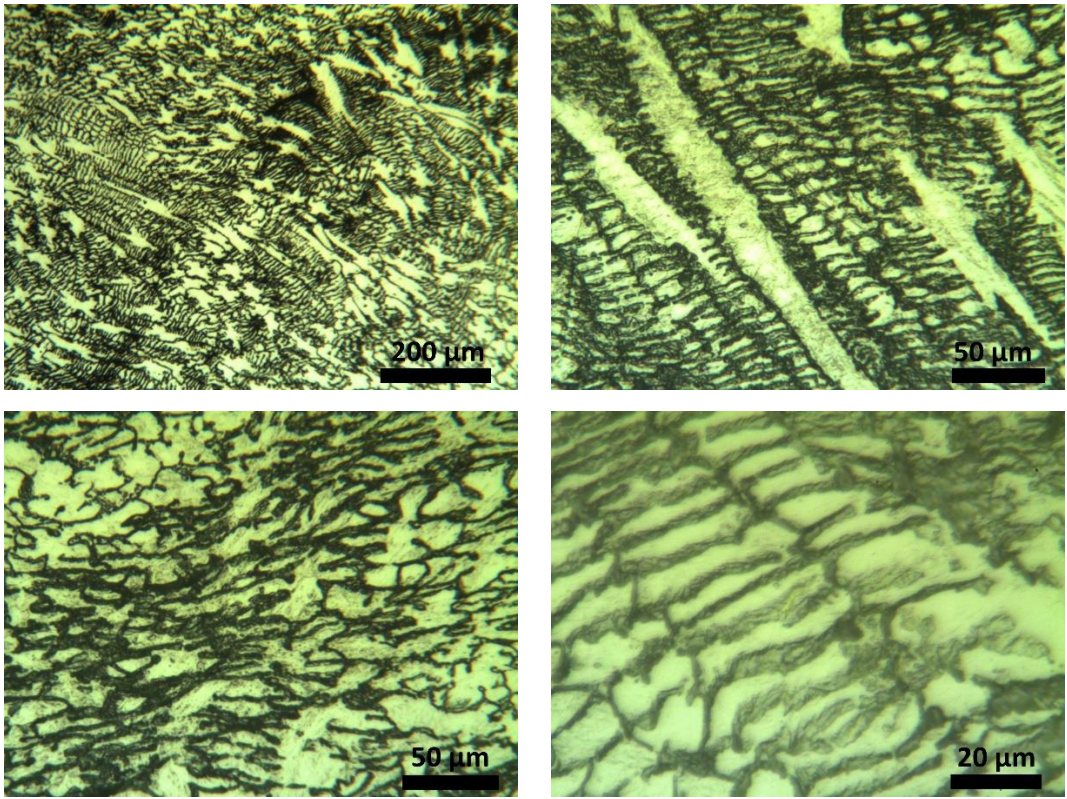

Figure 5. Observation of highly deformed areas on the investigated surface after compression test

Because the shape of dendritic regions (bright regions) in Figures 3 and 4 is relatively similar to the shape of the dendritic regions in the as cast condition, microstructures in Figures 3 and 4 may be attributed to small amounts of plastic deformation. As a result, yielding points observed at small strains (Figure 1), may be attributed to the microstructures in Figures 3 and 4. Furthermore, because new phases can be seen in Figure 4, therefore it may be concluded that the yield points at small strains in Figure 1 might be due to a phase transformation. It should be noted that no phase transformation was observed inside of some distorted dendrites (Figure 3). So, other deformation mechanisms may also be active during the early stages of deformation. Song et al. investigated the triple yielding behavior of metastable $Cu_{47.5}Zr_{47.5}Al_5$ composites [10]. Some of the deformation mechanisms reported by Song et al. [10] may also be relevant here: 1) The constraint effect of the harder phase (BCC/B2), 2) The pre-existence of small amount of martensitic phases and 3) The initiation and development of martensitic transformation. Because BCC/B2 phases are usually considered to be not deformable, the constraint effect of the B2/BCC phases may be very important here. If BCC/B2 phases can introduce constraints for the deformation of dendrites, then by changing the amount and shape of B2/BCC phases, the mechanical behavior of alloy should change. Further studies are needed to investigate these points.

In order to accurately assess the mechanical behavior observed in Figure 1, one needs to investigate the microstructure at certain amounts of strain. Simple approaches could be proposed in this regard. One approach could be producing a sample with a gradient of strain by preforming the rolling on a wedge sample. This is schematically shown in Figure 6. Another approach could be preforming the compression test on a conic sample, or a simpler approach may be performing a hardness test which will cause a gradient of strain around the indentation point. These experiments may be used for more accurately investigating the deformation behavior of alloy introduced here, or in general for investigating the deformation behavior of alloys with stress/strain-induced martensitic phase transformation (SIM).

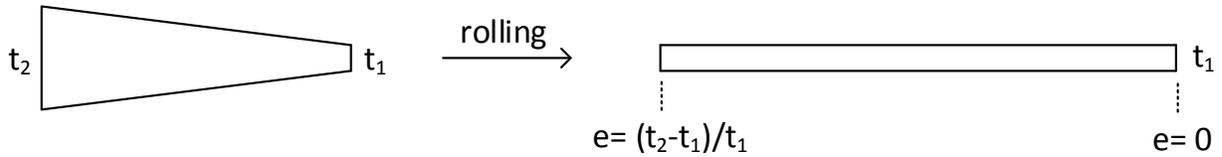

Figure 6. Performing rolling on a wedge sample for producing a gradient of strain along the sample

In summary, the compressive mechanical behavior of alloy $Al_{14}Co_{41}Cr_{16}Fe_{11}Ni_{18}$ is reported in this work. Three yielding points were observed during the compression test. According to the microstructural studies, the yielding points at small strains may be due to a stress/strain-induced martensitic phase transformation and the constraints of the harder phase during deformation.